\documentclass[prb, fleqn,twocolumn, showpacs, showkeys]{revtex4}% Physical Review B

\newcommand{\vo}{\mbox{VO$_2$}}
\newcommand{\voo}{\mbox{V$_2$O$_3$}}
\newcommand{\vooo}{\mbox{V$_2$O$_5$}}
\usepackage{amsmath}
\usepackage{graphicx, amssymb}
\usepackage[dvips]{epsfig}
\usepackage{etex}
\begin{document}

\title{Role of the V-V dimerization in insulator-metal transition and optical transmittance of pure and doped $\vo$ thin films}
\author{S. S. Majid$^{1}$, S. R. Sahu$^{2}$, A. Ahad$^{1}$, K. Dey$^{2}$, K. Gautam$^{2}$, F. Rahman$^{1}$, P. Behera$^{2}$, U. Deshpande$^{2}$, V. G. Sathe$^{2}$, D. K. Shukla$^{2, *}$} 
\affiliation{$^1$ Department of Physics, Aligarh Muslim University, Aligarh-202002, India\\ 
$^2$ UGC-DAE Consortium for Scientific Research, Indore-452001, India}
\email{dkshukla@csr.res.in}

\date{\today}

\begin{abstract}
Insulator to metal (IMT) transition (T$_t$ $\sim$ 341 K) in the $\vo$ accompanies transition from an infrared (IR) transparent to IR opaque phase. Tailoring of the IMT and associated IR switching behavior can offer potential thermochromic applications. Here we report on effects of the W and the Tb doping on the IMT and associated structural, electronic structure and optical properties of the $\vo$ thin film. Our results show that the W doping significantly lowers IMT temperature ($\sim$ 292 K to $\sim$ 247 K for 1.3\% W to 3.7\% W) by stabilizing the metallic rutile, $\it{R}$, phase while Tb doping does not alter the IMT temperature much and retains the insulating monoclinic, $\it{M1}$, phase at room temperature. It is observed that the W doping albeit significantly reduces the IR switching temperature but is detrimental to the solar modulation ability, contrary to the Tb doping effects where higher IR switching temperature and solar modulation ability is observed. The IMT behavior, electrical conductivity and IR switching behavior in the W and the Tb doped thin films are found to be directly associated with the spectral changes in the V 3$\it{d_{\|}}$ states.
\end{abstract}

 \maketitle

\section{Introduction}
Vanadium dioxide ($\vo$) is known to undergo first order insulator to metal transition (IMT) from the low temperature monoclinic insulating phase $\it{(P2_1/c)}$ to the high temperature rutile metallic $\it{(P4_2/mnm)}$ phase around the transition temperature T$_t$ $\sim$ 341 K. Structural transformation of $\vo$ from the rutile, $\it{R}$, structure to the monoclinic, $\it{M1}$ structure transforms the equally spaced V atoms, oriented along the rutile $\it{c}$ axis, into paired (dimerized) and tilted V atoms in the monoclinic structure $\cite{Zyl75,Ima98,Adl68}$. The IMT in $\vo$ is accompanied by dramatic modification in optical properties like switching in transmittance/reflectance of the infrared radiations with the insulating phase being transparent to the IR while the metallic phase blocks them. This property of $\vo$ is potentially useful for designing the thermochromic smart windows which can serve as the best energy management devices as these windows switch the solar heat gain $\cite{Zho13,Wu13,Zha14}$. 
\par Changes in the IR and visible transmittance of the $\vo$ thin films across the IMT can be explained in terms of changes in the band structures, explained in context of the Goodenough model $\cite{Goo71,Goo73}$. According to this model, in the rutile metallic phase of the $\vo$ the octahedral crystal field splits the V 3$\it{d}$ degenerate orbitals into doubly degenerate $e_g^{\sigma}$ and triply degenerate $t_{2g}$orbitals. The $t_{2g}$ orbitals are further splitted into the doubly degenerate $e_g^{\pi}$ orbitals and the single $\it{d_{x^2-y^2}}$ (d$_{\|}$) orbital. Both the $e_g^{\pi}$ and $e_g^{\sigma}$ states hybridize with O 2$\it{p}$ orbitals to form anti-bonding $\pi^{\ast}$ and $\sigma^{\ast}$ molecular orbitals while the d$_{\|}$ orbitals are oriented along the rutile $\it{c}$- axis. In the insulating phase pairing and the tilting of the Vanadium atoms along the rutile $\it{c}$- axis results in splitting of the d$_{\|}$ orbitals into filled bonding and empty anti-bonding orbitals. Band gap of $\sim$ 0.65 eV in the $\vo$ insulating phase arises due to energy difference between the bonding V 3$\it{d_{\|}}$ and the $\pi^{\ast}$ bands while in the metallic phase energy gap vanishes due to overlapping of the non bonding 3$\it{d_{\|}}$ and the $\pi^{\ast}$ bands at the Fermi level. Decrease in the IR transmittance of the $\vo$ with increase in the temperature is directly associated with vanishing of the $\vo$ insulating band gap ($\sim$ 0.65 eV) and increase in the free carrier concentration in the high temperature metallic phase. Optical transitions in the visible spectral range occur between lower O 2$\it{p}$ valence states to the $\pi^{\ast}$ conduction bands $\cite{Die17,Gav72}$. 

 \par For practical applications of the $\vo$ material as thermochromic smart windows, the IMT temperature of the $\vo$ requires to be tailored down to the room temperature (RT). Chemical doping of electron rich valence shell metal ions viz W$^{6+}$, Mo$^{6+}$ and Nb$^{6+}$ and optimizing the amount of strain in epitaxial $\vo$ thin films have been found effective ways to reduce the T$_t$ of the $\vo$ to the room temperature $\cite{Bat11,Pat12,Tan12,Die17,Mur02}$. Doping of electron doped ions Tungsten (W$^{6+}$), have been found effective in reducing the transition temperature to RT, while doping of the hole doped Terbium ions (Tb$^{3+}$), has not significant impact on the transition temperature lowering $\cite{Wan16}$. There is always a trade off between the transition temperature reduction, solar modulation ability and the luminous transmittance in the $\vo$ thin films. Normally an increase in the luminous transmittance is accompanied by the decrease in the solar modulation $\cite{Lin16,Bur02}$. W doping is found to reduce the solar modulation ability and the luminous transmittance $\cite{Bur02}$ while Tb doping enhances the luminous transmittance accompanied with the decrease in the solar modulation ability $\cite{Wan16}$. Two systems (W and Tb doped $\vo$) show contrast effect on the electrical and optical properties of the $\vo$ thin films. It is essential to completely study both the W and the Tb $\vo$ systems and explore the role of W and Tb in modifying the $\vo$ IMT and optical properties.
\par We have synthesized the pure $\vo$ and W (1.3\%, 2.9\% and 3.7\%) \& Tb (1.3\%, 3.3\% and 4.6\%) doped $\vo$ thin films on the quartz substrates using the pulsed laser deposition technique. By combining structural, electrical, optical and electronic structure studies we present an insightful of the effects of the W and Tb doping on the IMT and optical properties of the $\vo$. Our results directly show the doping induced alterations to the V-V dimerization in the $\vo$ thin film, on which electrical and optical properties are largely dependent.

\section{Experimental details}
KrF excimer laser ($\lambda$ = 248 nm, repetition rate of 5Hz and pulse energy of 210 mJ) was focused onto a target (pressed pure and W \& Tb doped $\vooo$) with a fluence of 1.1 mJ/cm$^2$. Ultrasonically cleaned quartz substrates were maintained at a temperature of 700$^{\circ}$C during the deposition. All depositions were performed in an oxygen partial pressure of 10 m Torr. X-ray diffraction (XRD) was carried out with a Bruker D8 x-ray diffractometer with Cu K$\alpha$ radiation. Temperature dependent resistivity measurements were performed in the standard four point configuration. Temperature dependent Raman spectra were collected in backscattering geometry using a 10 mW Ar (475 nm) laser as an excitation source coupled with a Labram-HRF micro-Raman spectrometer equipped with a 50X objective. The X-ray photoelectron spectroscopy (XPS) measurements were performed using an Omicron energy analyzer (EA-125) with a Al K$_\alpha$ (1486.6 eV) X-ray source. Soft X-ray absorption spectroscopy (SXAS) across the V $\it{L_{3,2}}$ and the O $\it{K}$ edges were carried out in the total electron yield (TEY) mode at the beam line BL-01, Indus-2 at RRCAT, Indore, India. Energy resolution during SXAS measurements at the oxygen $\it{K}$ edge energy was $\sim$ 250 meV. Bulk $\vo$ was measured for reference. V-K edge X-ray absorption near edge spectroscopy (XANES) measurements were performed at beamline P64, PETRA III, DESY, Hamburg, Germany in fluorescence mode. Optical transmittance spectra were collected using UV-VIS-NIR spectrometer, Perkin Elmer USA Model: Lambda 950, operating in the spectral range of 175 to 3300 nm having resolution $\sim$ 0.02 nm (UV-VIS) and $\sim$ 0.2 nm (NIR). Doping \% used throughout the paper are calculated using the XPS spectra of the Tb and the W.

\begin{figure}[hbt]
\centering
\includegraphics[width=0.5\textwidth]{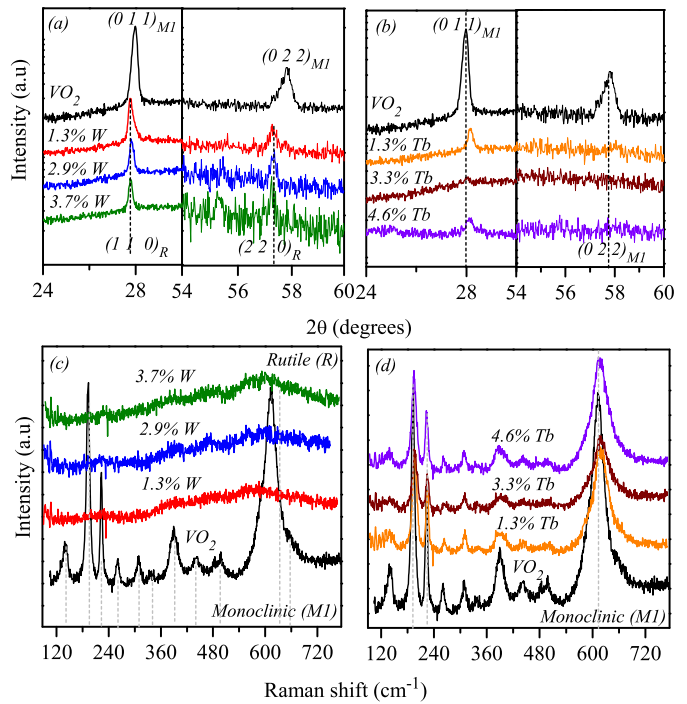}
\caption{Room temperature XRD patterns of (a) the pure and the W doped $\vo$ thin films, (b) the pure and the Tb doped $\vo$ thin films. Room temperature Raman spectra of (c) the pure and the W doped $\vo$ thin films and (d) the pure and the Tb doped $\vo$ thin films.}
\label{XRD}
\end{figure}
\begin{figure}[hbt]
 %\centering
\includegraphics[width= 0.5\textwidth]{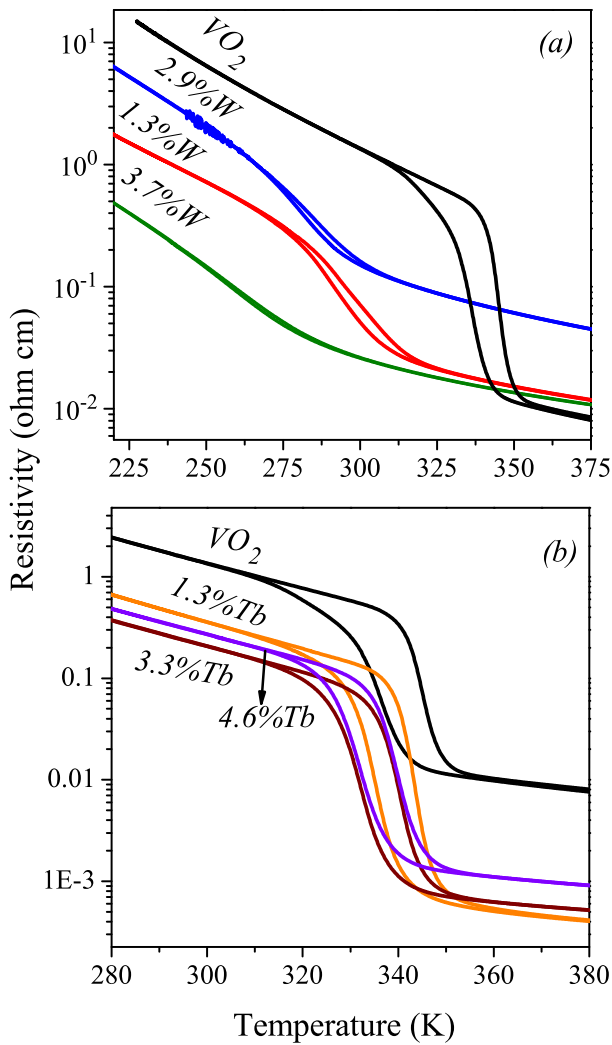}
\caption{Temperature dependent resistivity of (a) the pure and the W doped $\vo$ thin films (b) the pure and the Tb doped $\vo$ thin films.}
\label{T_resistance} 
\end{figure}

\section{Results and Discussions} 
Figure ~\ref{XRD} (a) shows X-ray diffraction (XRD) of the pure and  W doped thin films. XRD confirms growth of the $\vo$ monoclinic ($\it{M1}$) phase in the undoped thin film, visualized from the diffraction peaks observed at 2$\theta$ $\sim$ 27.97 $^{\circ}$ and 57.79 $^{\circ}$ corresponding to the reflections from the $\vo$ monoclinic (0 1 1) and (0 2 2) planes respectively $\cite{Maj18,Sep08}$. Tungsten doped $\vo$ thin films are grown in the rutile ($\it{R}$) phase, evident from 2$\theta$ positions of the diffraction peaks at $\sim$ 27.78 $^{\circ}$ and 57.27 $^{\circ}$, which correspond to reflections from the rutile (1 1 0) and (2 2 0) planes, respectively $\cite{Sep08}$. Further confirmation about grown phases in the pure and the W doped $\vo$ thin films comes from the Raman spectra shown in the Figure ~\ref{XRD} (c). Pure $\vo$ thin film shows seven A$_g$ and three B$_g$ Raman modes at $\sim$ 139, 192, 222, 309, 337, 389, 613 cm$^{-1}$ and 260, 441, 496, 659 cm$^{-1}$, respectively, which belongs to the $\vo$ monoclinic, M1, phase $\cite{Shi17}$. The A$_g$ symmetry Raman modes $\omega_{v_1}$ and $\omega_{v_2}$ at $\sim$ 192 and 222 cm$^{-1}$ are assigned to the V-V vibrations while rest of the observed Raman peaks are related to V-O vibrations $\cite{Ma08,Che11}$. Complete disappearance of the $\it{M1}$ vibrational modes observed in all the W doped thin films confirm growth of the rutile phase, consistent with earlier reports $\cite{Maj18, Yan14}$. XRD patterns of the Tb doped thin films shown in the Figure ~\ref{XRD} (b) exhibit slight decrease in crystalline nature and shifting of the monoclinic, $\it{M1}$, (0 1 1) diffraction peak towards higher 2$\theta$, compared to the pure $\vo$ thin film. Shifting of the diffraction peaks to higher 2$\theta$ values with increase in the Tb doping \% is attributed to decrease in unit cell size of the $\vo$. Raman spectra shown in the Figure ~\ref{XRD} (d) further confirms growth of the monoclinic, $\it{M1}$, phase in the Tb doped thin films and substantiates to XRD results. In contrast to the W doping and no phase transformation is visualized due to the Tb doping.

\begin{table*}[t]
\centering 
\caption{Parameters T$_t$, $T_{sc}$, $T_{oc}$ and IMT strength obtained from the temperature dependent resistivity, Raman (cooling cycle) and transmittance spectroscopy (cooling cycle) in the pure and the doped $\vo$ thin films. The $\it{d}$$_{\|}$ intensity in the pure and the doped $\vo$ thin films is manifested from the XAS measurements}
\begin{tabular} {c c c c c c }\hline 
\textbf{sample} &\textbf{T$_t$ (K)} &\textbf{$T_{sc}$ (K)} &\textbf{$T_{oc}$ (K)} &\textbf{IMT strength} &\textbf{$\it{d}$$_{\|}$ intensity} \\
& \textbf{resistivity} & \textbf{Raman} & \textbf{Transmittance} & \textbf{resistivity} & \textbf{XAS}  \\
 \hline   
 \textbf{{VO$_2$}} & \textbf{340.1} & \textbf{322.3}  & \textbf{320.7} & \textbf{56.6} & \textbf{0.55} \\
[0.5 ex] 
 $\textbf{1.3\% W}$ & \textbf{292.2} & \textbf{241.8}  & \textbf{285.2} & \textbf{21.4} & \textbf{0.42} \\
[0.5 ex] 
 $\textbf{2.9\% W}$ & \textbf{275.8} & \textbf{198.9}  & \textbf{246.2} & \textbf{18.8} & \textbf{0.45} \\
[0.5 ex] 
 $\textbf{3.7\% W}$ & \textbf{247.3} & \textbf{122.5}  & \textbf{217.1}  & \textbf{11.1} & \textbf{0.46} \\ 
[0.5 ex]
 $\textbf{1.3\% Tb}$ & \textbf{339.2} & \textbf{329.3}  & \textbf{341.6} & \textbf{271.4} & \textbf{0.64} \\ 
[0.5 ex]
 $\textbf{3.3\% Tb}$ & \textbf{335.7} & \textbf{328.2}  & \textbf{338.5} & \textbf{142.5} & \textbf{0.62} \\ 
[0.5 ex]
 $\textbf{4.6\% Tb}$ & \textbf{335.5} & \textbf{322.3}  & \textbf{330.5} & \textbf{109.9} & \textbf{0.66} \\ 
  \hline
\end{tabular}
\label{TB}%       
\end{table*}

\par Temperature dependent resistivity ($\rho$) measurements of the pure and the W/Tb doped $\vo$ thin films are shown in the Figures ~\ref{T_resistance} (a) and ~\ref{T_resistance} (b). The characteristic IMT is observed in all the thin films. To infer quantitative information from temperature dependent resistivity measurements, the IMT in all the thin films were fitted by the Boltzmann function ($\it{f(T)}$ = 1 - $\frac{1}{1+e^{\frac{T-T_{0}}{dT}}}$ equation (1)). $\it{f}$ is the resistivity, ($\rho$), T$_0$ is the transition temperature and dT signifies the transition width. IMT temperature (T$_t$) is defined as an average of transition temperatures measured in the cooling and the heating cycles. The IMT temperature for the pure and the doped $\vo$ thin films are illustrated in the Table I. Tungsten doping has effectively reduced the IMT transition temperature of the $\vo$ thin film. In the $\sim$ 1.3\% W doped $\vo$ thin film IMT is observed near at room temperature ($\sim$ 292 K), which is consistent with the earlier reports $\cite{Tak12,Shi10}$. Tb doping has small impact on the IMT temperature of the $\vo$ thin film compared to the W doping. Resistivity measurements are also used to calculate the IMT strength defined as resistivity ratio $\rho$ (T$_t$-$\Delta$)/$\rho$ (T$_t$+$\Delta$). Calculated values of the IMT strength of the pure and the doped $\vo$ thin films are listed in the Table I. 

\par
\begin{figure}[hbt]
 \centering
\includegraphics[width= 0.5\textwidth]{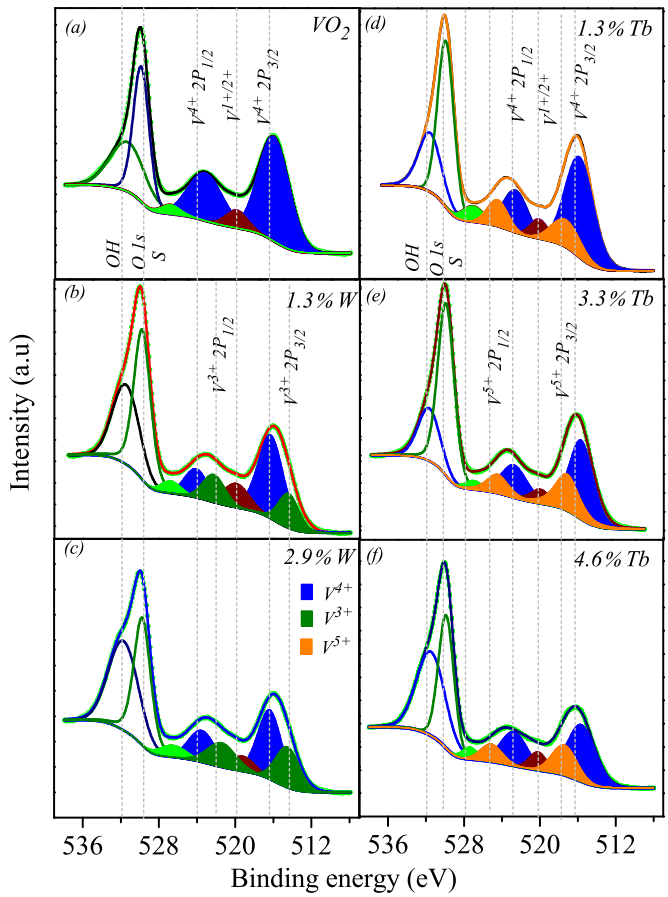}
\caption{O 1$\it{s}$ and V 2$\it{p}$ XPS of (a) the pure $\vo$ (b) 1.3\% W (c) 2.9\% W (d) 1.3\% Tb, (e) 3.3\% Tb and (f) 4.6\% Tb doped $\vo$ thin films}
\label{XPS}
\end{figure}

\begin{figure}[hbt]
 \centering
\includegraphics[width= 0.5\textwidth]{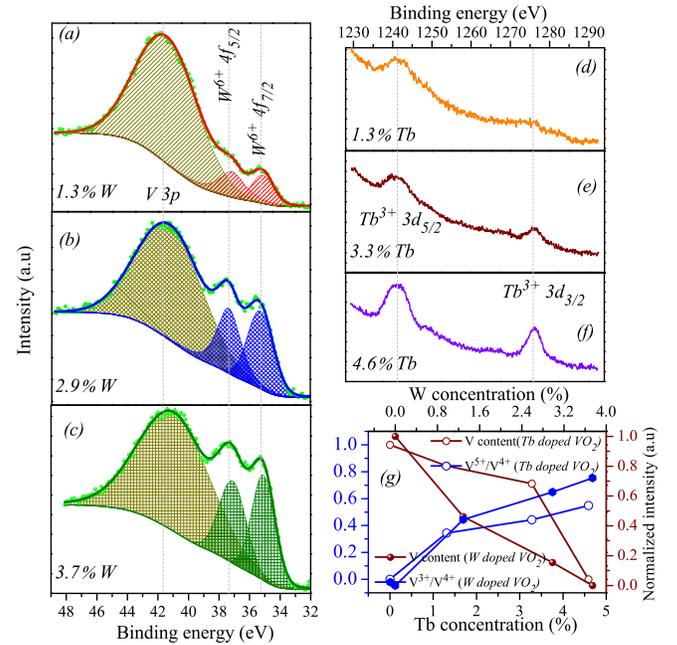}
\caption{(a, b and c) W 4$\it{f}$ XPS of the 1.3\%, 2.9\% and 3.7\% W doped $\vo$ thin films. (d, e and f) Tb 3$\it{d}$ XPS of the 1.3\%, 3.3\% and 4.6\% Tb doped $\vo$ thin films. (g) Variation of the $V^{5+}$/$V^{4+}$ and $V^{3+}$/$V^{4+}$ ratios for the Tb and W doping concentrations, along with the doping dependent Vanadium content variation in the $\vo$ thin film.}
\label{XPS_Tb_W}
\end{figure}

 For correct estimation of chemical states of the Vanadium and the dopant elements we have carried out XPS measurements, shown in the Figure ~\ref{XPS}. XPS of the pure $\vo$ thin film (Figure ~\ref{XPS} (a)) confirms the +4 oxidation state of the Vanadium, which is visible from the binding energy positions of the spin-orbit split features V 2$\it{p_{3/2}}$ $\sim$ 516 eV and V 2$\it{p_{1/2}}$ $\sim$ 523 eV. The features at $\sim$ 531.3 eV, $\sim$ 530 eV and $\sim$ 527 eV correspond to the OH concentration, O 1$\it{s}$ and the V 2$\it{p}$$_{3/2}$ satellite peak, respectively $\cite{Zim98,Gre04}$. The spectral feature at $\sim$ 520 eV is assigned to V$^{1+/2+}$ oxidation state which may arises due to reduction of some of the V$^{4+}$ due to the oxygen deficiency. W doping (Figures ~\ref{XPS} (b and c))  induces the spectral features at binding energies $\sim$ 514 eV and $\sim$ 522 eV, belonging to $V^{3+}$ 2$\it{p}$$_{3/2}$ and 2$\it{p}$$_{1/2}$ states, while Tb (Figures ~\ref{XPS} (d, e and f)) doping results in the inclusion of $V^{5+}$ 2$\it{p}$$_{3/2}$ and $V^{5+}$ 2$\it{p}$$_{1/2}$ states at the binding energy values $\sim$ 517 eV and $\sim$ 524 eV. 
XPS spectra shown in the Figures ~\ref{XPS_Tb_W} (a-c) confirm the +6 oxidation state of the W which is evident from the binding energy positions of 4$\it{f}$$_{7/2}$ $\sim$ 35 eV and 4$\it{f}$$_{5/2}$ $\sim$ 37 eV features, while the Figures ~\ref{XPS_Tb_W} (d-f) represent +3 oxidation state of the Tb which is visualized from the spectral features at $\sim$ 1241 eV and $\sim$ 1276 eV belonging to Tb 3$\it{d}$$_{5/2}$ and 3$\it{f}$$_{3/2}$ states. The feature at $\sim$ 41.5 eV is due to V 3$\it{p}$ states $\cite{Pan17,Zha12}$. In order to maintain charge neutrality in the W and the Tb doped $\vo$ thin films, some Vanadium atoms change their oxidation state to +3 and +5 to compensate the higher ($W^{6+}$) and the lower (Tb$^{3+}$) charge states. Figure ~\ref{XPS_Tb_W} (g) shows decrease in the Vanadium content with increase in the doping percentage of the W and the Tb atoms and the decrease in the Vanadium content is found to be accompanied by an increase in the ratio of $V^{3+}$/$V^{4+}$ and $V^{5+}$/$V^{4+}$ for W and Tb doped $\vo$, respectively. 
 
\par
\begin{figure}[hbt]
 \centering
\includegraphics[width= 0.5\textwidth]{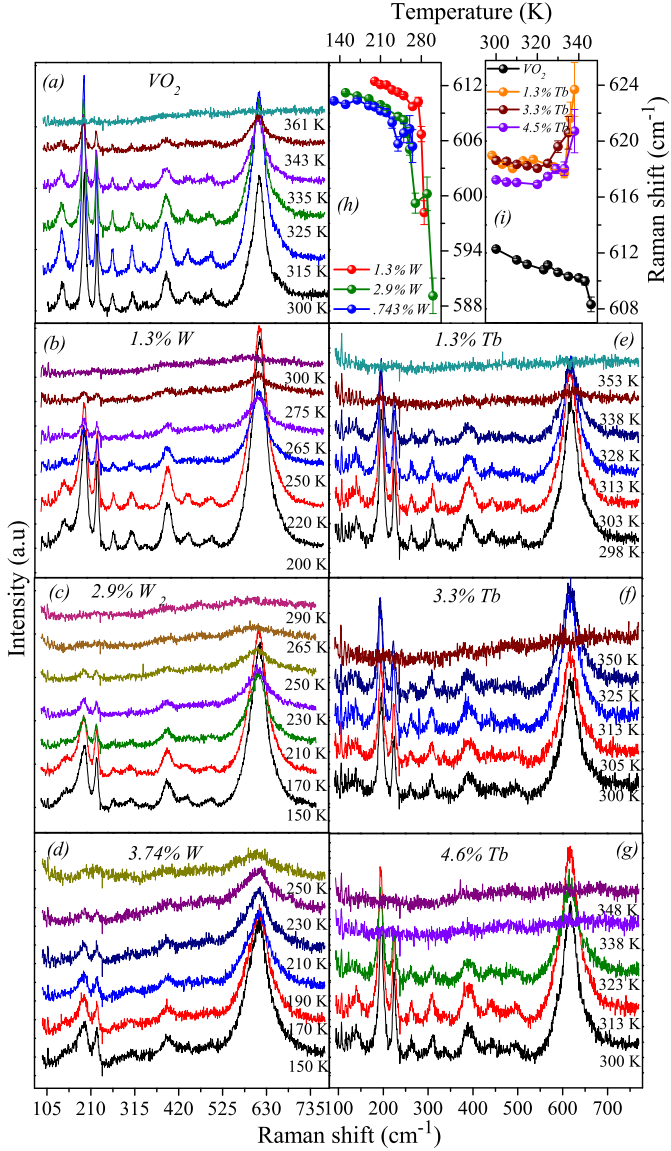}
\caption{(a-g) Temperature dependent Raman spectra of the pure and the W/Tb doped $\vo$ thin films collected in the cooling cycle.  Temperature dependent frequency shift of the $\omega_{0}$ Raman mode positions obtained from Lorentz fitting in (h) W doped (i) pure and the Tb doped $\vo$ thin films.}
\label{T_Raman}
\end{figure}

 Temperature dependent Raman spectra of the pure and the W/Tb doped $\vo$ thin films are shown in the Figures ~\ref{T_Raman} (a-g), measured in the cooling cycles. Temperature dependent structural transition form the low temperature monoclinic ($\it{M1}$) structure to the high temperature rutile ($\it{R}$) structure is observed in all the thin films without presence of any intermediate structures ($\it{M2}$ and $\it{T}$) which are usually accessible via chemical doping in the $\vo$ $\cite{Maj18}$. Temperature dependent frequency position of the Raman mode $\omega_{0}$, of the pure and the W/Tb doped $\vo$ thin films are plotted in the Figures ~\ref{T_Raman} (h) and (i). There is not any anomalous change observed in the temperature dependent frequency position of the Raman mode, $\omega_{0}$. Softening of the Raman mode positions due to lattice thermal expansion is accompanied by large and sudden change at the structural transition temperature. In the Tb doped $\vo$ thin films frequency shift of the $\omega_{0}$ mode from $\sim$ 613 cm$^{-1}$ (observed in the Pure $\vo$ thin film, to $\sim$ 619 cm$^{-1}$) reflects significant influence of the Tb doping on the V-O vibrational mode compared to the W doping, where the $\omega_{0}$ mode stays around $\sim$ 613 cm$^{-1}$. Normally the higher frequency shift of the $\omega_{0}$ mode position in the pure $\vo$ thin film  is considered as fingerprint for identification of the $\vo$ $\it{M2}$ and $\it{T}$ structures $\cite{Maj18,Atk12,Ma08}$, however in the Tb doped $\vo$ thin films higher frequency shift of the $\omega_{0}$ mode to $\sim$ 619 cm$^{-1}$ is much less compared to that observed for the $\it{T}$ phase ($\sim$ 629 cm$^{-1}$) and for the $\it{M2}$ phase ($\sim$ 642 cm$^{-1}$).

\par
\begin{figure}[hbt]
 \centering
\includegraphics[width= 0.5\textwidth]{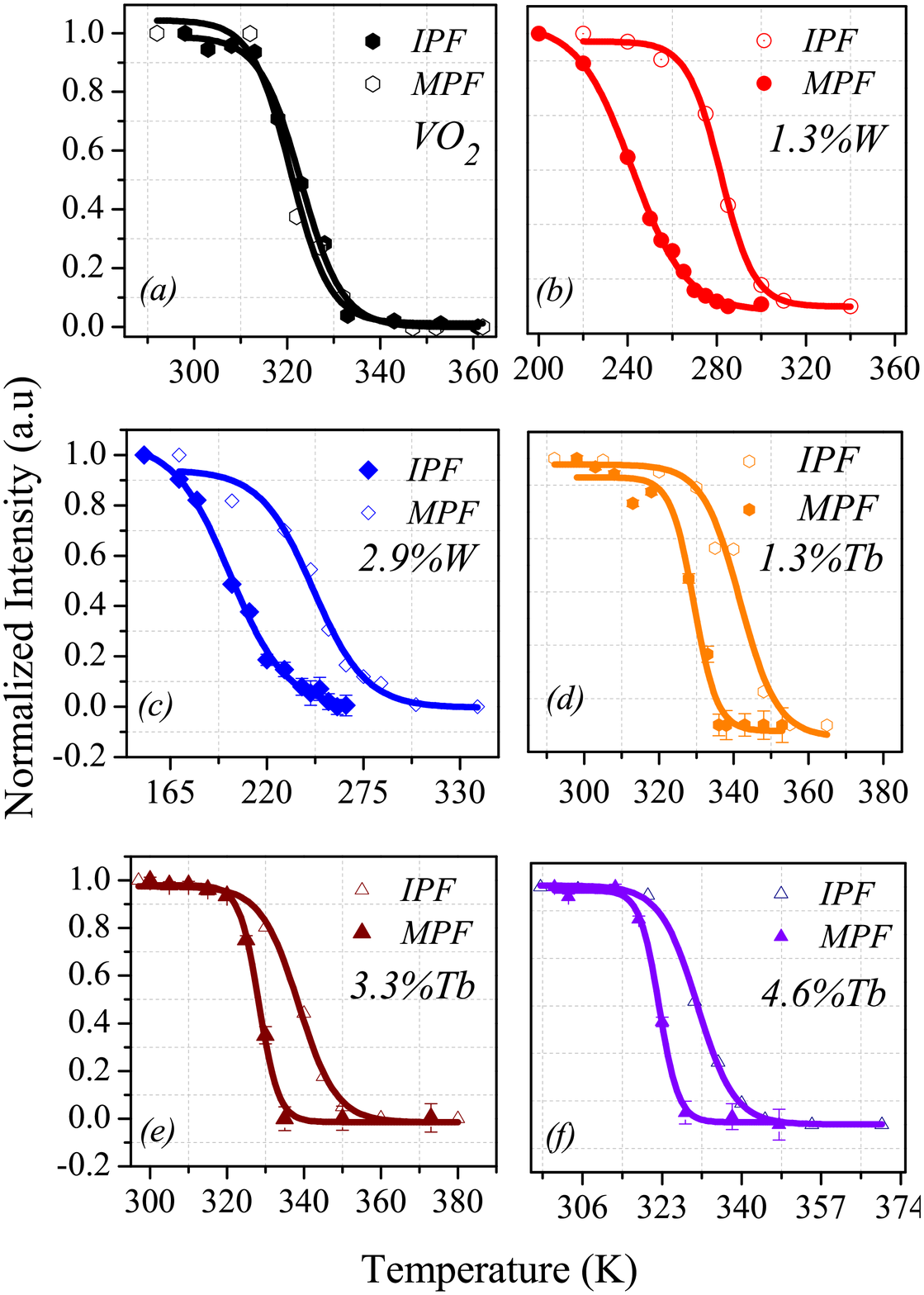}
\caption{(a-f) Temperature dependent variation of the insulating phase fraction (IPF) obtained from optical transmittance measurements, and monoclinic phase fraction (MPF) obtained through the $\omega_{0}$ Raman mode in the pure and the W/Tb doped $\vo$ thin films. Data shown in these graphs are from the cooling cycles.}
\label{IPF_MPF}
\end{figure}

\begin{figure}[hbt]
 \centering
\includegraphics[width= 0.5\textwidth]{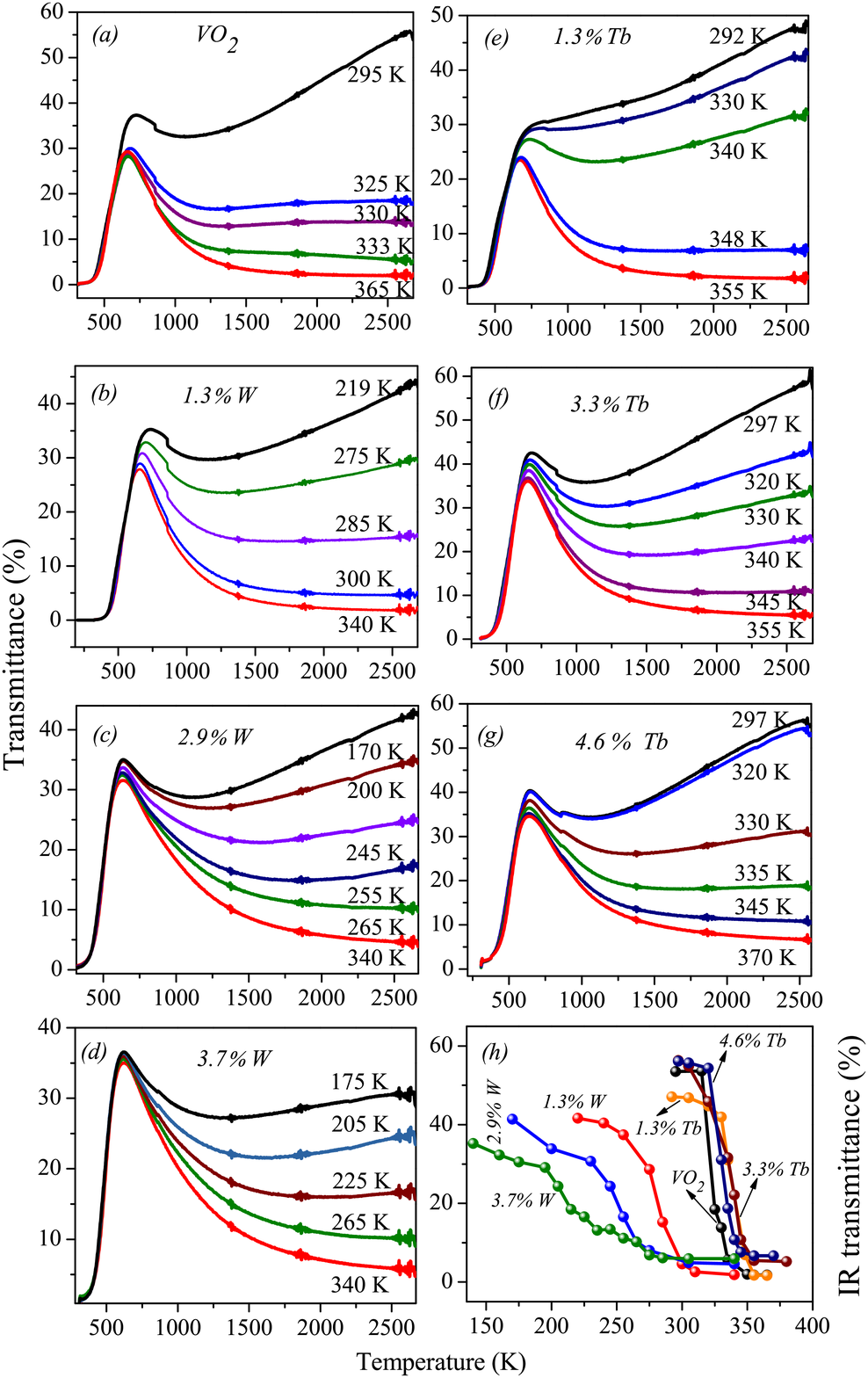}
\caption{(a-g) Temperature dependent optical transmittance spectra of the pure and the W/Tb doped $\vo$ thin films. (h) Temperature dependent IR transmittance at $\sim$ 2500 nm of the pure and the W/Tb doped $\vo$ thin films in the cooling cycles.}
\label{T_transmittance}
\end{figure}

 Recently, Yoav $\textsl{ et al.,}$\cite{Kal19} used XRD and IR spectroscopic techniques to understand coupling between the IMT and the structural phase transition in the $\voo$ thin films and discussed the non reliability of the resistance measurements in determination of the correct insulating and the metallic phase fractions. We have extracted the the monoclinic phase fraction (MPF) and the insulating phase fraction (IPF) from temperature dependent Raman and optical transmittance spectra (discussed later, see Figure ~\ref{T_transmittance}), respectively. Following formula has been used to calculate the IPF (in the range, 2000 $cm^{-1}$ to 2500 $cm^{-1}$).

\begin{equation}
IPF(T) = \frac{T_r (T)-T_{rM}}{T_{rI}-T_{rM}}\;
\end{equation}   
where $\it{T_{rI}}$ and $\it{T_{rM}}$ are transmittance of the pure insulating and the pure metallic phases and $\it{T_r (T)}$ is the transmittance at temperature, T. The calculated IPF of the pure and the W/Tb doped $\vo$ thin films are compared with the monoclinic phase fraction (MPF) obtained from temperature dependent integrated intensity of the $\omega_{0}$ Raman mode in the Figures ~\ref{IPF_MPF} (a to f). Calculated transition temperatures, T$_{oc}$ and T$_{sc}$ from the optical transmittance measurement and the Raman spectroscopy data using the equation (1), are listed in the Table I. In the pure $\vo$ thin film IMT is found to be coupled with the structural transition while W and Tb doping are found to stabilize the rutile structure prior to the metallic state. Impact of the W doping in the rutile structure stabilization is much more significant compared to the Tb doping.    

\par For thermochromic applications optical properties of the pure and the doped $\vo$ thin films are characterized using temperature dependent UV-VIS-NIR transmittance spectra shown in the Figure ~\ref{T_transmittance}. The spectral transmittance in the wavelength range of $\sim$ 250 nm to $\sim$ 2500 nm of the pure (Figure ~\ref{T_transmittance} (a)), W doped (Figures ~\ref{T_transmittance} (b-d)) and the Tb doped (Figures ~\ref{T_transmittance} (e-g)) $\vo$ thin films are used to calculate the solar transmittance,$\it{T_{sol}}$. For the luminous transmittance, $\it{T_{lum}}$, calculations spectral range $\sim$ 382 nm to $\sim$ 732 nm is used.  Both the solar and the luminous transmittance are calculated using following formula:
\begin{equation}
T_{sol/lum} = \frac{\int \phi_{sol/lum} (\lambda)T_r(\lambda) d\lambda}{\int \phi_{sol/lum}(\lambda)d\lambda}
%T_{sol/lum} = \frac{\int \phi_{sol/lum} (\lambda)T_r(\lambda) d\lambda}{\int \phi_{sol/lum}(\lambda)d\lambda}}\;
\end{equation}    
where T$_r$($\lambda$) is the spectral transmittance at wavelength $\lambda$. $\phi$$_{sol}$ ($\lambda$) is the solar spectral irradiance for the sun standing 37$^{\circ}$ above the horizon (for the air mass 1.5) and the $\phi$$_{lum}$ is the standard luminous efficiency function for the photopic vision of human eyes $\cite{Kan10,Zha14}$. Using the integrated solar and the luminous transmittance, we can calculate the solar and the luminous modulation of the pure and the doped $\vo$ thin films using following formulas:
\begin{equation}
\Delta T_{sol} = T_{r(sol, I)} - T_{r(sol, M)}\;
\end{equation}  
\begin{equation}
\Delta T_{lum} = T_{r(lum, I)} - T_{r(lum, M)}\;
\end{equation}
 
where T$_{r(sol, I)}$, T$_{r(sol, M)}$, T$_{r(lum, I)}$ and T$_{r(lum, M)}$ are the solar and the luminous transmittance of the insulating and the metallic states, respectively. Parameters obtained from transmittance spectra of the pure and the W/Tb doped $\vo$ thin films are illustrated in the Table II. IR switching in the pure and the W/Tb doped $\vo$ thin films can be visualized from the Figure ~\ref{T_transmittance} (h) where temperature dependent change in the IR transmittance at the wavelength, $\lambda$ $\sim$ 2500 nm is plotted, in the cooling cycle. The IR switching transition temperatures for the pure and the W/Tb doped $\vo$ thin films are calculated from the IR transmittance Vs T curve, measured in the cooling cycle, using the equation (1) and is found equivalent to $T_{oc}$ (see Table I). The near room temperature T$_{oc}$, T$_{r(lum)}$ $\geq$ 40\% and the $\Delta$$\it{T_{sol}}$ $\geq$ 10\% are considered as the best criteria for the $\vo$ thin films to be employed in thermochromic windows applications $\cite{Li12,Zha14}$. From the Table II it is found that the W doping in the $\vo$ thin film effectively reduces the IR switching temperature to near room temperature $\sim$ 282 K (measured in the cooling cycle) but is accompanied with decrease in the solar modulation ability, consistent with earlier reports $\cite{Gra07,Lin16}$. With the Tb doping IR switching occurs at higher T$_{oc}$, while the solar modulation ability is maintained around $\sim$ 15\% which is much large compared to the earlier report on Tb doped $\vo$ thin films $\cite{Wan16}$. There is always a trade off between the solar modulation ability and the luminous transmittance in the $\vo$ thin films i.e., an increase in the luminous transmittance is accompanied by the decrease in the solar modulation $\cite{Lin16,Bur02}$. However, in the Tb doped $\vo$ thin films increase in the luminous transmittance is accompanied with the good and stable solar modulation ability (see Table II). 
\begin{figure}[hbt]
 \centering
\includegraphics[width= 0.4\textwidth]{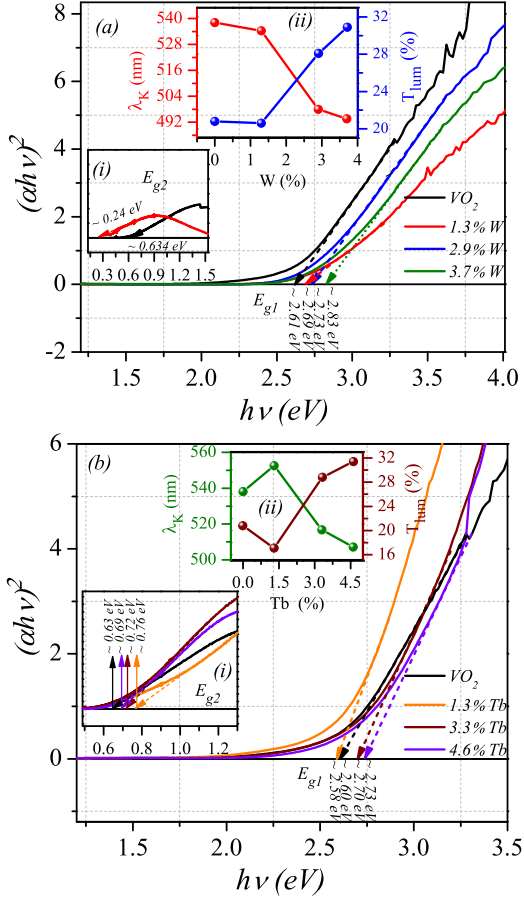}
\caption{(a) Tauc plot (($\alpha$ $\it{h}$$\nu$)$^2$ Vs $\it{h}$$\nu$) of the pure and the W doped $\vo$ thin films at room temperature. Insets: ($\it{i}$) Tauc plot in the IR range of photon energies ($\it{ii}$) variation in the T$_{lum}$ and the $\lambda$$_K$ with the W doping percentage. (b) Tauc plot of the pure and the Tb doped $\vo$ thin films at room temperature. Insets: ($\it{i}$) Tauc plot in the IR range of photon energies ($\it{ii}$) variation in the T$_{lum}$ and the $\lambda$$_K$ with the Tb doping percentage.}
\label{Absorbance}
\end{figure}    
\par In order to understand the transmittance observations it is essential to calculate band gap of the pure and the W/Tb doped $\vo$ thin films corresponding to the visible and the IR transparency, which can be extracted from the Tauc plot shown in the Figures ~\ref{Absorbance} (a) and (b). Tauc plot is graphical description between ($\alpha$$\it{h}$$\nu$)$^n$ Vs $\it{h}$$\nu$ obtained from the Tauc's equation $\cite{Tau72}$ given as:
\begin{equation}
(\alpha h\nu)^n = A (h\nu - E_g)\;
\end{equation}     
 where n = 1/2, 1/3, 2 and 2/3 for the indirect allowed, indirect forbidden, direct allowed and direct forbidden optical transition, respectively, $\alpha$ is absorption coefficient obtained from the transmittance spectra, A is a constant and $\it{h}$$\nu$ is the photon energy. Usually, in the optical experiments direct band gap is visualized because of the lower excitation probability among different symmetry points during the optical transition $\cite{Liu10}$.The band gap is obtained by extrapolating the linear portion of the curve while y ordinate is zero. Two types of band gaps are observed i.e., $E_{g1}$ and $E_{g2}$. $E_{g2}$ corresponds to energy gap between the V 3$\it{d_{\|}}$ and $\pi^{\ast}$ bands while the $E_{g1}$ is due to the optical transition between the O 2$\it{p}$ and $\pi^{\ast}$ bands $\cite{Wan17,Shu13}$ (Figures ~\ref{Absorbance} (a) and (b)). The IR switching temperature in all the thin films is found to depend on the energy band gap $E_{g2}$. The lowering of the IR switching temperature in the W doped $\vo$ thin films (see Table II) is attributed to decrease in the energy band gap, $E_{g2}$, from $\sim$ 0.63 eV in the pure to $\sim$ 0.24 eV in the 1.3\% W doped $\vo$ thin film, while in the Tb doped $\vo$ thin films we have observed highest IR switching temperature $\sim$ 341.6 K in the $\sim$ 1.3\% Tb doped $\vo$ thin film having the largest band gap $E_{g2}$ $\sim$ 0.74 eV. 

\par The band gap $E_{g1}$ corresponds to the luminous transparency of the thin films, higher the $E_{g1}$ value higher will be the visible transparency $\cite{Wan17,Shu13,Zh13}$. Inset ($\it{ii}$) of the Figures ~\ref{Absorbance} (a) and (b) show the variation of the luminous transmittance along with the absorption edge ($\lambda$$_K$), obtained from the derivative of the transmittance spectra. The blue shift in the absorption edge along with an increase in the luminous transmittance in the $\vo$ thin film due to W and Tb doping are found to be consistent with an increase in the energy band gap values $E_{g1}$.   

\begin{table*}[t]
\centering 
\caption{Parameters obtained from the temperature dependent optical transmittance measurements of the pure and the W/Tb doped $\vo$ thin films}
\begin{tabular} {c c c c c c c c c}\hline 
$\textbf{sample}$ &\textbf{T$_{lum,I}$} &\textbf{T$_{lum,M}$} &\textbf{$\Delta$T$_{lum}$} &\textbf{T$_{sol,I}$} &\textbf{T$_{sol,M}$} &\textbf{$\Delta$T$_{sol}$} \\
 \hline   
 $\textbf{{VO$_2$}}$ & \textbf{20.8} & \textbf{19.5}  & \textbf{1.3} & \textbf{32.4} & \textbf{17.0} & \textbf{15.4} \\
[0.5 ex] 
 $\textbf{1.3\% W}$ & \textbf{20.6} & \textbf{19.4}  & \textbf{1.2} & \textbf{30.2} & \textbf{16.4} & \textbf{13.7} \\
[0.5 ex]  
 $\textbf{2.9\% W}$ & \textbf{28.1} & \textbf{26.4}  & \textbf{1.67} & \textbf{32.7} & \textbf{23.4} & \textbf{9.2} \\
[0.5 ex]  
 $\textbf{3.7\% W}$ & \textbf{30.9} & \textbf{30.8}  & \textbf{0.07} & \textbf{32.6} & \textbf{26.6} & \textbf{5.9} \\
[0.5 ex]  
 $\textbf{1.3\% Tb}$ & \textbf{17.1} & \textbf{14.9}  & \textbf{2.2} & \textbf{28.9} & \textbf{13.6} & \textbf{15.2}\\
[0.5 ex]  
 $\textbf{3.3\% Tb}$ & \textbf{28.8} & \textbf{26.1}  & \textbf{2.6} & \textbf{38.9} & \textbf{23.7} & \textbf{15.1} \\
[0.5 ex]  
 $\textbf{4.6\% Tb}$ & \textbf{31.4} & \textbf{27.9}  & \textbf{3.5} & \textbf{38.8} & \textbf{25.4} & \textbf{13.4}   
 \\ \hline
\end{tabular}
\label{TB}%      
\end{table*}

\par
\begin{figure}[hbt]
 \centering
\includegraphics[width= 0.5\textwidth]{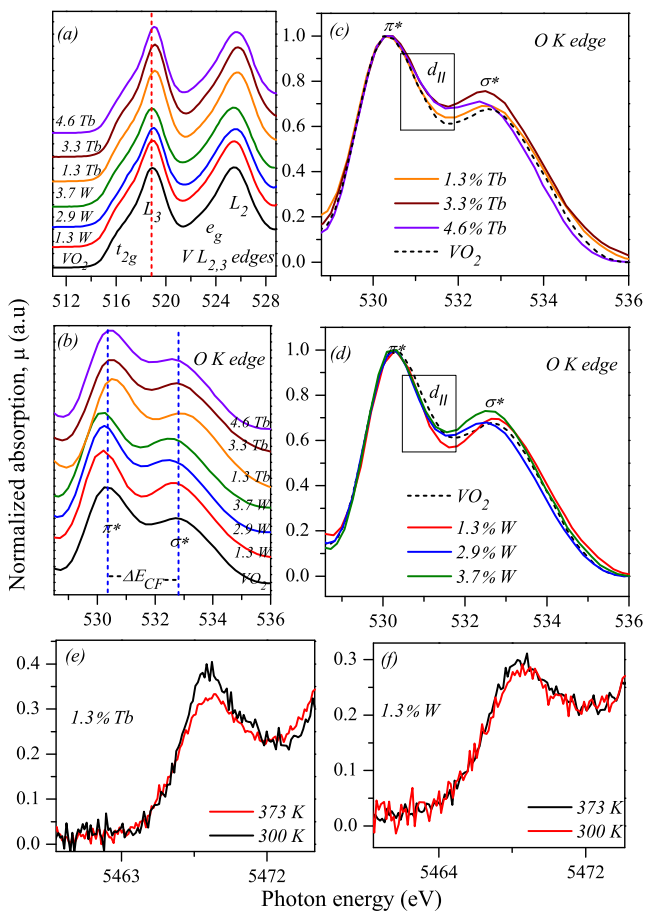}
\caption{Room temperature (a) V L$_{2,3}$ (b) O K edge XAS measurements of the pure and the W/Tb doped $\vo$ thin films. Normalized O K edge XAS of (c) the pure and the Tb doped (d) the pure and the W doped $\vo$ thin films. The V $\it{K}$ pre edge XAS (e) of the 1.3\% Tb doped and (f) of the 1.3\% W doped $\vo$ thin films at 300 K and 373 K.}
\label{XAS}
\end{figure}

 In the following we will discuss about origin of doping induced changes in the electrical conductivity and optical properties of the $\vo$. The V $\it{L_{2,3}}$ edge X-ray absorption spectra of the pure and the W/Tb doped $\vo$ thin films are shown in the Figure ~\ref{XAS} (a). The vanadium $\it{L_{2,3}}$ spectrum arises due to the transition from the V 2$\it{p}$ core level (spin orbit splitted V 2$\it{p_{3/2}}$ and 2$\it{p_{1/2}}$) to the unoccupied V 3$\it{d}$ states and consists of two pronounced features at $\sim$ 518.8 eV (L$_3$) and $\sim$ 525.5 eV (L$_2$). The features at $\sim$ 516.0 eV and $\sim$ 522.6 eV are due to the transitions to crystal field splitted t$_{2g}$ part of the V 3$\it{d}$ states $\cite{Hav05,Zaa85}$. The energy positions of the $\it{L_{3}}$ $\sim$ 518.8 eV and $\it{L_{2}}$ $\sim$ 525.5 eV features confirm 4+ oxidation state of the Vanadium. With the W and Tb doping in the $\vo$ thin film the slight shift in the energy positions of the $\it{L_{3}}$ and $\it{L_{2}}$ states is observed which is due to the additional V$^{3+}$ and V$^{4+}$ states created by the W and Tb doping, as inferred from the XPS measurements also. 

The O $\it{K}$ edge spectrum, see the Figure ~\ref{XAS} (b), results from transition of the O 1$\it{s}$ electrons to the vacant O 2$\it{p}$ states which are hybridized with the V 3$\it{d}$ orbitals and form the conduction band. Peaks at $\sim$ 530.3 eV and $\sim$ 532.8 eV are due to transitions into the $\pi^{\ast}$ and the $\sigma^{\ast}$ hybridized bands $\cite{Ruz07,Goo73}$. Doping of the W has clearly resulted in shift of the $\pi^{\ast}$ band towards the lower energies as compared to the pure $\vo$ thin film which signify the stabilization of the metallic state. The Tb doping shifts the $\pi^{\ast}$ band towards higher energy signifying the stabilization of the insulating state. The crystal field splitting energy ($\Delta$E$_{CF}$) which is the energy separation between the $\pi^{\ast}$ and the $\sigma^{\ast}$ states is found to vary with the W and the Tb doping. Variations in the crystal field splitting energy values due to the W and the Tb doping in the $\vo$ thin film is also susceptible to different extent of V-V interactions. In order to clearly visualize the spectral changes among the XAS spectra of pure and doped $\vo$ thin films we have carried out the normalization and horizontal shift of the $\pi^{\ast}$ states to $\sim$ 532.8 eV as shown in the Figures ~\ref{XAS} (c and d). The asymmetry observed in the $\pi^{\ast}$ bands around $\sim$ 531 eV, which is marked in both the Figures \ref{XAS} (c and d), is assigned to the unoccupied d$_{\|}$ states $\cite{Ruz08,Yeo15}$ and a decrease and an increase in the intensity of the d$_{\|}$ feature is observed in the W and Tb doping, respectively. For better visualization and comparison pure $\vo$ spectra is shown in dotted line. The integrated intensity of the d$_{\|}$ states for the pure and the doped $\vo$ thin films is noted in the Table 1. The $\pi^{\ast}$orbitals in the $\vo$ thin film point in between the ligands and have less orbital overlap with the O 2$\it{p}$ orbitals, while the $\sigma^{\ast}$ orbitals are directed toward the ligand having a higher overlap with the O 2$\it{p}$ orbitals. Therefore, spectral weight of the $\pi^{\ast}$ band is highly sensitive to the V-V interactions while spectral weight of the $\sigma^{\ast}$ band is related to the direct V-O interactions $\cite{Ruz08}$. The pre-edge structures at $\sim$ 5468.3 eV from the V $\it{K}$ edge XAS observed in the 1.3\% Tb and the W doped $\vo$ thin films at the temperatures 330 K and 373 K are shown in the Figures \ref{XAS} (e) and (f). The pre-edge peak intensity is highly sensitive to the local geometrical structure around the absorbing atom (V) and increases directly with increase in distortion of the regular octahedral VO$_6$ units $\cite{Maj18}$. The $\vo$ rutile, $\it{R}$ phase has less distorted octahedron compared to the monoclinic, $\it{M1}$ phase, hence an increase in intensity in the 300 K spectra compared to the 370 K spectra of the Tb doped $\vo$ thin film represent the transition from the $\it{R}$ phase to the $\it{M1}$ phase while no such signature is observed in the W doped $\vo$ thin film indicating stable rutile phase at the room temperature and 373 K. By combining the resistivity and XAS measurements we find that the IMT strength of the pure and the W/Tb doped $\vo$ thin films is directly related to the unoccupied spectral weight of the V 3$\it{d_{\|}}$ states. W doping is found to decrease the IMT strength while Tb doping increases the IMT strength which is accompanied by decrease and an increase in the spectral weight of the V 3$\it{d_{\|}}$ states (see the Table I and the Figure \ref{XAS} (c)). 

\begin{figure}
 \centering
\includegraphics[width= 0.5\textwidth]{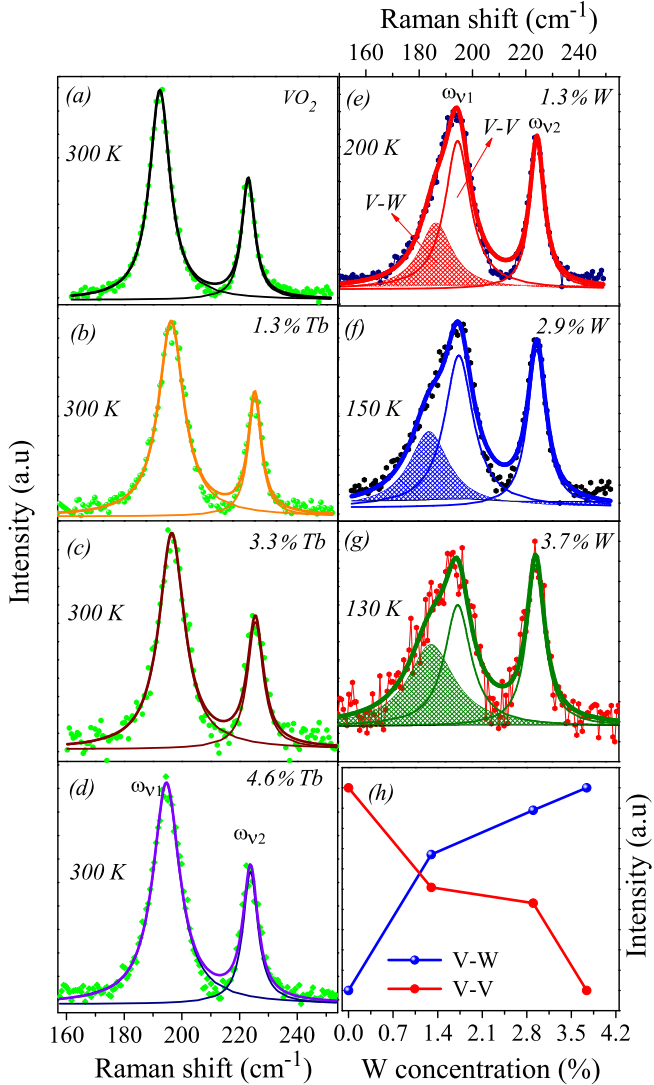}
\caption{(a-d) Room temperature (300 K) Raman integrated intensity of the $\omega_{v_1}$ and $\omega_{v_2}$ vibrational modes in the pure and the Tb doped $\vo$ thin films. (e-g) Raman integrated intensity of the $\omega_{v_1}$ and $\omega_{v_2}$ vibrational modes in the 1.3\%, 2.9\% and 3.7\% W doped $\vo$ thin films at 200 K, 150 K and 130 K, respectively. (h) Raman intensity variation of the V-O and V-W modes with the increase in the W\% concentration.}
\label{I_Raman}
\end{figure}  

\par Spectral weight of the $\it{d_{\|}}$ states observed in the XAS measurements is directly related to the Vanadium-Vanadium dimerization along the rutile $\it{c}$ axis $\cite{Koe06,Qua16}$. The V-V dimerization in the $\vo$ can be directly visualized from the $\omega_{v_1}$ and the $\omega_{v_2}$ vibrational modes in Raman spectra shown in the Figures \ref{I_Raman} (a to g) as both these modes are assigned to V-V vibrations and the $\omega_{v_1}$ frequency mode gives the twisting vibration of the Vanadium dimers along the rutile $\it{c}$ axis. Raman mode $\omega_{v_1}$ in the pure and the Tb doped $\vo$ thin films is fitted with a single Lorentzian while in the W doped $\vo$ thin films vibrational mode is fitted with two Lorentzian peaks at $\sim$ 186 and 195 cm$^{-1}$. The additional peak at the lower frequency value $\sim$ 186 cm$^{-1}$ is assigned to the V-W vibrational mode as effective mass of the V-W system is larger as compared to the V-V system. Presence of the V-W vibrational modes in the W doped $\vo$ thin films signify replacement of the V atoms by the doped W atoms along the $\vo$ rutile $\it{c}$ axis, hence the loss of V-V dimerization. On contrary in the Tb doped $\vo$ thin film absence of any such extra mode signify that the Tb is not doped along the rutile $\it{c}$ axis therefore there is no loss of V-V dimerization. Figure ~\ref{I_Raman} (h) shows the intensity variation of the V-V and the V-W vibrational modes in the W doped sample. Increase in intensity of the V-W vibrational mode and corresponding decrease in intensity of  the V-V vibrational mode with increase in the W doping can be visualized. According to Tang $\textsl{ et al.,}$\cite{Tan85} electron dopant W atoms break the V$^{4+}$-V$^{4+}$ dimers and result in formation of the V$^{3+}$-W$^{6+}$ and V$^{3+}$-V$^{4+}$ pairs along the rutile $\it{c}$ axis due to electron transfer from the W to the V ions. The electronic charge transfer from the W atoms resulting in formation of V$^{3+}$ ions (see Figure ~\ref{XPS}) stabilizes the metallic state and the loss of the Vanadium dimerizations favors the rutile symmetry $\cite{Tan12,Tan85}$. Band filling due to electron dopant W atoms results in stabilization of the metallic phase by melting down the electron-electron correlations, U, $\cite{Tak12}$. 
 \begin{figure}
 \centering
\includegraphics[width= 0.5\textwidth]{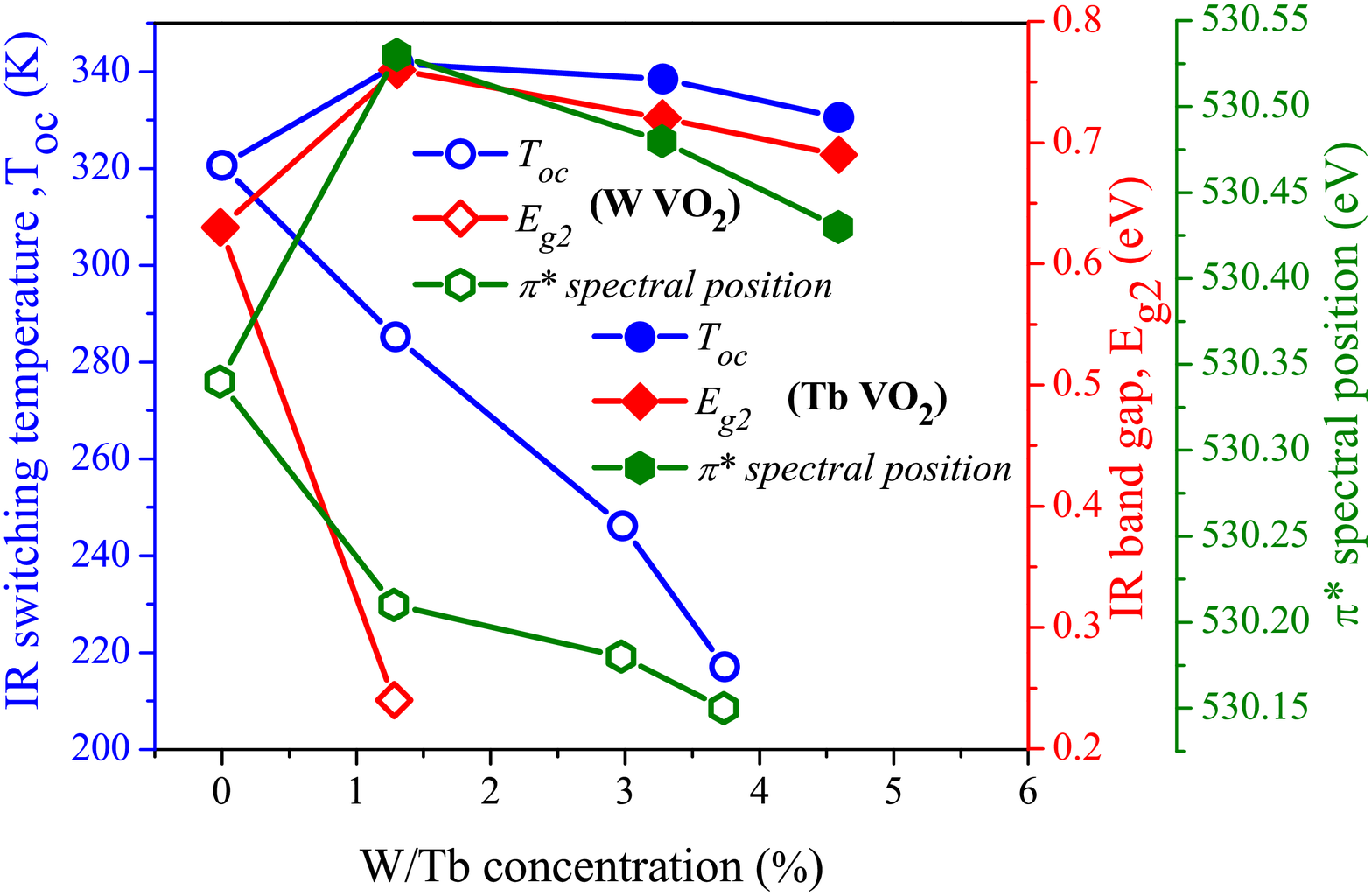}
\caption{Variation in the spectral position of the $\pi^{\ast}$ states, IR band gap (E$_{g2}$) and the IR switching temperature (T$_{oc}$) with the W\% and the Tb\% concentration in the $\vo$ thin films.}
\label{Discussion}
\end{figure}  
\par XAS and optical transmittance results reveal that the IR switching temperature, T$_{oc}$ is strongly related to the spectral weight and position of the V 3$\it{d_{\|}}$ and $\pi^{\ast}$ states (see Figure ~\ref{XAS}). Infrared switching temperature (T$_{oc}$) in all thin films, which is sensitive to the energy gap E$_{g2}$ (energy gap between the $\it{d_{\|}}$ and the $\pi^{\ast}$ states) is found to be directly related to the spectral position of the $\pi^{\ast}$ states (see Figure ~\ref{Discussion}). The spectral weight and position of the $\pi^{\ast}$ band is directly sensitive to the V-V interactions and reflects the extent of the V-V hybridization $\cite{Shr10}$. Larger V-V hybridization will stabilize the $\pi$ bonding states and simultaneously shift the $\pi^{\ast}$ anti-bonding states towards the higher photon energy. The decreasing T$_{oc}$ \& E$_{g2}$ with increasing W doping concentration and increasing T$_{oc}$ \& E$_{g2}$ with increasing Tb concentration are found to be concomitant with shift of the $\pi^{\ast}$ states spectral weight towards lower and higher energies, respectively. The solar modulation ability of all the thin films can be directly associated with their IMT strength. In the W doped $\vo$ thin films $\Delta$$\it{T_{sol}}$ decreases with decrease in the IMT strength. In the Tb doped $\vo$ thin films an increased and the stabilized $\Delta$$\it{T_{sol}}$ values (contrary to previous report $\cite{Wan16}$) can be attributed to an increased IMT strength observed from the resistivity measurements. Hence the solar modulation ability, $\Delta$$\it{T_{sol}}$, is found to be largely dependent on stabilization of the $\vo$ monoclinic, $\it{M1}$ phase and therefore on preservation of the V-V dimers.                              

\section{Conclusions}
In summary, changes in electrical, structural, optical and electronic structure properties of the pure and the W/Tb doped $\vo$ thin films have been studied. Electron dopant W atoms are found to significantly reduce the IMT temperature in the $\vo$ thin film by stabilizing the rutile metallic phase, contrary to the hole dopant Tb atoms which favor and stabilize the insulating monoclinic phase. Loss of the V-V dimerization is clearly observed due to the W doping while the Tb doping is found to strengthen the V-V dimerization. Loss of the V-V dimerization has been found to directly influence the spectral weight of the V 3$\it{d_{\|}}$ states, a decrease and an increase in spectral weight of the $\it{d_{\|}}$ states has been observed due to the W and the Tb doping, respectively. Changes in strength of the IMT, the electrical conductivity, the IR switching temperature and the solar modulation ability due to the W and Tb doping are found directly related to the spectral changes in the $\it{d_{\|}}$ states. Conclusions which are directly related to thermochromic applications of doped $\vo$ thin films are as follows, i) A decrease in the IR switching temperature is observed due to the W doping while an increase in the IR switching temperature is visualized due to the Tb doping. Near room temperature IR switching temperature ($\sim$ 285 K, cooling cycle) is observed at an optimum 1.3\% W concentration. ii) An increase in the luminous transmittance is observed for both, the W and the Tb doping. Luminous transmittance of $\sim$ 30.9\% and 31.4\%  is observed for the highest 3.7\% W and 4.6\% Tb dopings. iii) An increased and stabilized solar modulation ability ($\sim$ 15.2) is observed with the 1.3\% Tb doping and a decrease in the solar modulation ability ($\sim$ 13.7) due to an optimum 1.3\% W doping. Our results also clearly indicate that the W and the Tb co-doping may be a potential candidate to get a combination of lowered IMT temperature, large $\Delta$$\it{T_{sol}}$ and high $\it{T_{lum}}$ essential for smart window applications.      

%In conclusion, electron dopant W atoms are found to significantly reduce the IMT temperature in the $\vo$ thin film temperature by stabilizing the rutile metallic phase, contrary to the hole dopant Tb atoms which favor and stabilize the insulating monoclinic phase. Changes in strength of the IMT and the electrical conductivity due to the W and Tb doping are found directly related to the spectral changes observed in the $\it{d_{\|}}$ states. An increased luminous transmittance accompanied, by decrease in the IR switching temperature due to the W and the large stable solar modulation ability due to the Tb is observed in the $\vo$ thin films. The IR switching temperature and the solar modulation ability in all the pure and the doped $\vo$ thin films are found to be directly associated with amount of the V-V dimerizations.                

\section{Acknowledgments}
Authors are thankful to Sharad Karwal Rakesh Sah and Ajay Rathore for help during XPS, XAS and Raman measurements, respectively. R.J. Choudhary is gratefully acknowledged for providing the PLD setup. E. Welter is gratefully acknowledged for his support during hard X-ray absorption spectroscopy measurements, performed under India-DESY collaboration. K.G. thanks, Council of Scientific and Industrial Research, New Delhi, India, for support in form of SRF. D.K.S. acknowledges support from SERB, India, in the form of an early-career research award (ECR/2017/000712).
\bibliography{Tb_W_VO2}
\end{document}